\documentclass[review,number,sort&compress]{elsarticle}
\usepackage{amssymb}
\usepackage{lineno}

\journal{Nuclear Instruments and Methods in Physics Research A}
\begin{document}

\begin{frontmatter}

\title{DC vacuum breakdown in an external magnetic field}
\author[IAP]{S. Lebedynskyi\corref{cor1}}
\ead{lebos@nas.gov.ua}
\author[IAP]{O. Karpenko}
\author[IAP]{R. Kholodov}
\author[IAP]{V. Baturin}
\author[IAP,CERN]{Ia.~Profatilova}
\author[CERN]{N. Shipman}
\author[CERN]{W. Wuensch}
\cortext[cor1]{Corresponding author}
\address[IAP]{Institute of Applied Physics, National Academy of Sciences of Ukraine, \\
58, Petropavlivska St., 40000 Sumy, Ukraine}
\address[CERN]{CERN, European Organization for Nuclear Research, 1211 Geneva, Switzerland}
\begin{abstract}
The subject of the present theoretical and experimental investigations is the effect of the external magnetic field induction on dark current and a possibility of breakdown. The generalization of the Fowler-Nordheim equation makes it possible to take into account the influence of a magnetic field parallel to the cathode surface on the field emission current. The reduction in the breakdown voltage due to the increment in electron-impact ionization was theoretical predicted. Experimentally shown that the presence of a magnetic field about a tenth as a large as the cutoff magnetic field \citep{Hull} reduces the breakdown voltage by $10\%$ to $20\%$ for practically all cathodes no matter what their surface treatment.

\end{abstract}


\end{frontmatter}


\section{Introduction}
Presently, experiments in particle physics require progressively higher energies, and thus, higher accelerating gradients. Earlier experiments have revealed high-vacuum breakdowns occurring due to energy input by an electromagnetic RF field, providing the accelerating rate of about $100~MV/m$  \citep{Kidemo}. Experiments with accelerating structures of a prototype compact linear collider (CLIC) showed  the advent of breakdowns when such gradients were tested. Therefore the acceleration could not be provided along the full length of the collider \citep{Wuensch}. And to achieve the required electron- and positron energies of $3~TeV$ did not seem feasible. As was pointed out in \citep{Tarasova,Slivkov1,Slivkov2,Latham,Ilic}, the first stage of breakdown is the electron field emission. A considerably increased electric field and higher emitted currents are observed on metal surfaces exhibiting asperities or some other irregularities. Phenomena accompanying the breakdown (heating, melting, evaporation of asperities with subsequent crater formation in the electrodes) are considered at length in \citep{Kyritsakis,Tian,Kyritsakis2,Palmer}.

Pointing to the fact that RF- and DC breakdowns have many features in common, the researchers focus attention on the underlying physics of DC breakdowns produced by a pulsed electric field \citep{Wuensch2}. The pulsed DC experiments are intended to explore possible ways of controlling the breakdown phenomenon, more specifically, the breakdown rate (BDR), with a view to reducing it. BDR is the ratio of the number of pulses with breakdown to the total number of pulses in the case of a constant electric field and a constant gap between the electrodes. The breakdown probability must not exceed $10^{-7}$ $1/m$ for CLIC to be effective and the accelerating gradient to reach $100~MV/m$. Since each breakdown results in beam losses and damage to accelerating structures, various ways of improving the structure resistance to high-gradient breakdowns are discussed, viz. vacuum deposition of a more refractory metal film onto the surface, ion implantation, surface conditioning with a smaller gradient, positioning of an accelerating structure in an external magnetic field (solenoid), etc.

In \citep{Blatt,Gogadze,Lebedynskyi, Lebedynskyi2, Lau, Hull} the authors analyze the magnetic field effect on the emission current. Blatt \citep{Blatt} theoretical examines a flat metal surface emitting electrons and external magnetic field normal to it. The effect of the field on the emitted current is determined by a modified spectrum of conduction electrons. Blatt supposed that the penetrability of potential barrier at the metal-vacuum interface is independent of the magnetic field, which supposition was borne out in \citep{Lebedynskyi, Lebedynskyi2}. According to the current density formula given in \citep{Blatt}, the emission current decreases in proportion to $B^2$. In addition to this decrease, there are also periodic current fluctuations. In \citep{Gogadze} a theoretical treatment is given to current fluctuations under field emission from a metal in a magnetic field normal to the specimen surface. The fluctuations were shown to be due to two reasons: fluctuations of the number of electron states in the magnetic field and fluctuations of the metal chemical potential.

Another theoretical study \citep{Lau} deals with the effect of a magnetic field parallel to the surface on the current. It is to be noted that electrons emitted from the metal surface would come back if the magnitude of the magnetic field induction is sufficient.
As a result, this may decrease the likelyhood of breakdown increasing the accelerating structure resistance. The cutoff magnetic field is given by the formula \citep{Hull}:
\begin{equation}
B_H=\sqrt{\frac{2mV}{ed^2}+\left({\frac{mv_0}{ed}}\right)^2},
\end{equation}
where m is the electron mass, $V$ is the voltage between the electrodes, $-e$ is the electron charge, $d$ is the interelectrode distance, and $v_0$ is the initial electron velocity. Using the accelerating gradient target of $100~MV/m$, we can calculate the limiting fields which for typical interelectrode distances of $10 \div 100~\mu m$ are $B_H=10.7\div 3.4~ T$. Thus, emission current  "switching-off" by means of an external magnetic field which is parallel to the metal surface is a possible way of preventing the high-vacuum high-gradient breakdowns.

So far no comprehensive theory has been proposed for the description of high-vacuum breakdown, only analyses of separate processes are currently available \citep{Palmer,Cherepnin}. A brief review of breakdown models without external magnetic fields suggests that the breakdown initiation can be best explained by ohmic heating due to emission current from surface asperities \citep{Palmer}. The processes of formation and evolution of the nano-tips on the metal surface were investigated in \citep{Jansson, Baibuz, Vigonski}. As is demonstrated in \citep{Stratakis}, electrons emitted by an electric field are accelerated and focused by a magnetic field on the other side of a cavity, heating its surface. This process can cause melting, vaporization and plasma formation that leads to breakdown. What is more, if the magnetic field is strong, the surface deformation may be substantial limiting eventually the accelerating gradient of the cavity. Field emission from a flat resonator in the presence of magnetic fields are discussed in \citep{Stratakis2}. It is shown that electrons emitted from the surface irregularities are focused by a magnetic field in small spots at some other location in the resonator heating the surface. It turns out that when the magnetic field is on the order of $T$, thermal stresses induced by a pulsed electron flow exceed the elastic limit, and the surface is prone to fatigue. In \citep{Palmer,Stratakis3, Stratakis4} the authors discuss a possibility to inhibit damage caused by electrons produced on the surface by designing RF cavities so that all high-gradient surfaces are parallel to an external magnetic field. Instead of focusing electrons the field would return them, with smaller energy, to their emission sites. In \citep{Stratakis5, Bowring} it was illustrated numerically and in\citep{Kochemirovskiy,Kochemirovskiy2} by experiment that using a magnetic field which is tangential to RF cavity walls, mechanical surface damage caused by field emission can be adequately suppressed, with the magnetic fields being on the order of the cutoff magnetic field $B_H$. Also, experiments were performed to examine the effect exerted on BDR by a relatively weak magnetic field (an order of magnitude weaker than the cutoff magnetic field, $B_H$) \citep{Shipman}. At present, however, no theoretical study is available providing a better understanding of the mechanisms underlying the breakdown occurrence.

This paper is a sequel to a previous publication \citep{Shipman}. The intent of the present work is to study theoretically and experimentally the effect of low (about a fraction of $B_H$) external magnetic fields on the dark current and breakdown rate.

\section{The analyze of the flows between electrodes}
The high-vacuum breakdowns are divided [3] into breakdowns at small gaps (tens of $\mu m$) and relatively small voltages $U<20~kV$ and large gaps and high voltages. In the case of small gaps, the following stages of vacuum breakdown are present: field emission of electrons from tips located on the surface of a cathode; field evaporation of the metal; heating of the tip, the appearance of the plasma near the cathode surface; microexplosion and breakdown (arc burning) \citep{Seznec}.

For large interelectrode distances, experimental results have shown that the breakdown voltage follows a power law regarding the electrode gap distance, with the superscript lying between 0.5 and 0.7 (in contradistinction to order 1 in the case of small gaps) \citep{Trump, Cranberg}.  But the general theory is not present now. Researchers \citep{Tarasova, Seznec} indicated the essential role of  processes in the gap. And an ionization of the residual and diffuse gases, adsorption, metal vaporization and bombarding electrodes by accelerated ions and electrons are among them. Therefore, it is necessary to consider in more detail the processes and currents flowing in the interelectrode gap to find the possibility of the magnetic field influence on the high-vacuum high-voltage breakdown.

Let us consider the constituents of the particle flows in the interelectrode gap before the breakdown. The entire flow can be divided into charged particles flow and neutral particles flow. The flow of charged particles (dark current) will be:
\begin{equation}
j_{dark}=j^{(-)}+j^{(+)}+j^{(+-)}+j^{(add)},
\end{equation}
where $j^{(-)}$ is the cathode current, $j^{(+)}$ is the anode current, $j^{(+-)}$ is the interelectrode (ionization) current, $j^{(add)}$ is an additional current that initiated under the influence of other external factors (cosmic radiation, etc.).

Neutral particles flow can be written as follows:
\begin{equation}
j^{0}=j_{rest}^{(0)}+j_{desor}^{(0)}+j_{diff}^{(0)}+j_{T,evap}^{(0)}+j_{Ion,atom}^{(0)},
\end{equation}
where $j_{rest}^{(0)}$ is the residual gas flow, $j_{desor}^{(0)}$ is the flow of desorbed atoms and molecules, $j_{diff}^{(0)}$ is the flow of diffuse atoms and molecules, $j_{T,evap}^{(0)}$ is the flow of the evaporated particles (due to local heating of the electrodes), $j_{Ion,atom}^{(0)}$ is the flow of particles knocked-out by ions falling on the electrodes. The first three terms can be reduced by improving the vacuum conditions and the cleaning of the cathodes and the vacuum chamber. This should be considering as a characteristic of the experimental installation. The last two terms arise only as a result of the flows between the electrodes.

Cathode current consists of field emission current\citep{Fowler} $j_{emis,e^-}^{(-)}$, the current of ions bombard the surface of the anode $j_{Ion,bombar}^{(-)}$ and the current of emitted ions from the cathode material $j_{emis,Me^-}^{(-)}$Then,
\begin{equation}
j^{(-)}=j_{emis,e^-}^{(-)}+j_{Ion,bombar}^{(-)}+j_{emis,Me^-}^{(-)}
\end{equation}
The anode current can be written as:
\begin{equation}
j^{(+)}=j_{emis,Me^+}^{(+)}+j_{e^-,bombar}^{(+)},
\end{equation}
where $j_{emis,Me^+}^{(+)}$ is the positive ions current, $j_{e^-,bombar}^{(+)}$ is the current of particles released due to bombardment of anode surface by emitted electrons.

The ionization current is composed of ionized metal atoms and molecules of gas (both impact ionization and field ionization of gas are possible):
\begin{equation}
j^{(+-)}=j_{e^-,Me^+}^{(+-)}j_{e^-,gas^+}^{(+-)}
\end{equation}

It is evident that in equations (2-6) the greatest value is the field electron emission current $j_{emis,e^-}^{(-)}$. But other terms can play a great role in the breakdown process due to the large mass of ions accelerated by a strong electric field. For example, Fowler-Nordheim equation which well describes field emission current \citep{Fowler} does not explicitly depend on kind of emitted particles. Therefore, we can assume that the electron current  from cathode and ion current from anode are equal. From the equality of the densities of field currents of electrons and ions follows a relation:
\begin{equation}
\frac{j_{Me^+}}{j_{e^-}}=\frac{v_{e^-}}{v_{Me^+}}=\sqrt{\frac{m_{e^-}}{m_{Me^+}}}\sim10^{-3}
\end{equation}
which coincides with \citep{Tarasova}. Despite the fact that the ion current is three orders less than the electron current it can lead to breakdown. Ion bombarding leads to significant destruction of the cathode surface with the possibility of further formation of an avalanche-like increase of the current in the interelectrode gap. Therefore, we can conclude that currents of different particles (not only electron current) in the interelectrode gap has a adverse influence on the breakdown resistant.

\section{The generalization of the Fowler-Nordheim equation influenced by a magnetic field}
Let's extend the Fowler-Nordheim equation \citep{Fowler} for the case of presence of an external magnetic field parallel to the metal surface. This can be done by writing the current density in a covariant four-dimensional form (in any other frame of reference that moves perpendicular to the direction of the electric field). Due to the Lorentz transformation for an electromagnetic field a magnetic field $\vec{B}\perp \vec{E}$ will appear in addition to the electric field. It is necessary to use two Lorentz invariants in this case:
\begin{equation}
\begin{array}{l}
I_1=F_{ik}F^{ik}\Rightarrow E^2-\left(cB\right)^2=inv \\
I_2=\varepsilon_{iklm}F^{ik}F^{lm}\Rightarrow E\cdot B=inv
\end{array}
\end{equation}
where $\varepsilon_{iklm}$ is an entirely antisymmetric unit tensor, $I_1$ is a true scalar, $I_2$ is a pseudoscalar (product $F^{ik}$ and its dual tensor). Since the pseudoscalar cannot enter into the true four-vector, only the first invariant $I_1$ remains.
We note that the character of the motion of a charged particle in perpendicular electric and magnetic fields is determined by the invariant $I_1$ \citep{Landau}.

In the case $I_1<0$ the external field is a magnetic type field. In this case a charged particle move in a trochoidal path and drift along the electrode surface.

In the case $I_1>0$ the external field is an electric type field. An electric field induce an acceleration of a charged particle and it moves to infinity with deflection from the axis under the influence of the magnetic field. Experimental values for study the magnetic field influence on the breakdown were chosen with respect to the formula (8).  The magnetic field was $0.48~T$ and the electric field was $144~ MV/m$ at the Large Electrodes System at CERN. Experiments at the Institute of Applied Physics, National Academy of Sciences of Ukraine (IAP NASU) were done with such fields: $E=100\frac{MV}{m}$ and $B=0.33~T$. Below we will assume that $I_1>0$ and use the corresponding field values.

The field emission current is theoretically well described by the Fowler-Nordheim equation \citep{Fowler}:
\begin{equation}
j_{FN}=A\cdot E^2\exp(-b/E),\\
\end{equation}
where
$$A=\frac{e}{2\pi h}\frac{\mu^2}{(\chi +\mu)\chi^\frac{1}{2}},~~~b=\frac{8\sqrt{2m}\pi \chi^\frac{3}{2}}{h}.$$
And were used follows symbols as $e$ is an electron charge, $\mu$ is the thermodynamic partial potential of an electron, $\chi$ is the thermionic work function, $h$ is a Planck constant.

We will use a covariance approach to estimate the effect of a magnetic field on the emission current density. We will be using the assumption that processes in a vacuum gap follows physics relations (in a covariant form) and are the equal in any inertial frame of reference.
The current density is a vector quantity. According to the differential form of the Ohm's law it reads:
\begin{equation}
\vec{j}=\sigma\vec{E},
\end{equation}
where $\sigma$ is a scalar. According to the Fowler-Nordheim equation (9) it equals:
\begin{equation}
\sigma=A\cdot E \exp(-b/E)
\end{equation}
It is necessary to replace $E$ in (9):
\begin{equation}
E\to\sqrt{E^2-\left(cB\right)^2}
\end{equation}
It's need to write the square of the electric field in front of the exponent in the expression for the field emission current density (9) in the form:
\begin{equation}
\sigma\sim \sqrt{I_1}\Rightarrow E^2\to \left|\vec{E}\right|\sqrt{I_1}
\end{equation}

Only the tensor of the electromagnetic field can be used by multiplying it to another vector characterizing the physical quantity in the process of field emission of electrons to obtain the vector of the electric field strength. Let's use for this a four-velocity of the new reference frame $u_k$:
\begin{equation}
u_k=\left(\gamma,\frac{\vec{V}}{c}\gamma \right),
\end{equation}
where $\gamma=\frac{1}{\sqrt{1-\frac{\upsilon^2 }{c^2}}}$

Denote the vector that is formed after multiplying the tensor by a four-velocity as $G^i=F^{ik}u_k$. The current density vector $\vec{j}$ is three components of the 4-vector $j^i$. $j^i$ can only be built by contracting the extra index $k$ of the tensor with $u_k$. Then, $j^i\sim F^{ik}u_k=G^i$. It is easy to see that this vector in the tensor form can be written as:
\begin{equation}
G^i=F^{ik}u_k=\left(\vec{E}\frac{\vec{V}}{c}\gamma,\vec{E}\gamma+\gamma\left[\vec{V}\times\vec{B}\right]\right)
\end{equation}

In the case of $B=0$ and taking into account that in this case the velocity of the new reference frame is $V=0$:
\begin{equation}
\vec{G}=\vec{E}
\end{equation}

Then the current density in the four-vector covariant takes the form:
\begin{equation}
j^i={A}F^{ik}u_k\cdot \sqrt{I_1}\cdot \exp{(-b/\sqrt{I_1})}
\end{equation}

Considering that $V=\frac{cB}{E}$, $\gamma$ takes form $\gamma=\frac{E}{\sqrt{E^2-c^2 B^2}}$. The absolute value of the spatial components of this expression gives the desired quantity of the current density taking into account the influence of the magnetic field ( $\vec{V} \|  \vec{B}$):
\begin{equation}
\left|\vec{j_B}\right|=A\cdot{E^2}\exp\left(-b/\sqrt{E^2-c^2B^2}\right)
\end{equation}

Equation (18) takes into account an external magnetic field perpendicular to the electric. In the absence of a magnetic field this equation is transformed into a well-known Fowler-Nordheim equation [27].

Fig.1 shows plots of field emission current obtained from equation (18) in Fowler-Nordheim coordinates. The influence of a magnetic field on the current density decrease with increase of an electric field strength.
\begin{figure}[h]
\center{\includegraphics[width=0.5\textwidth]{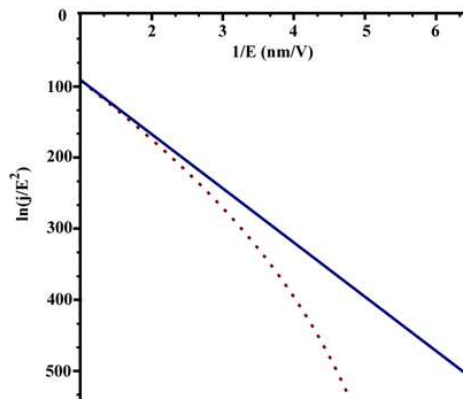}}
\caption{Influence of a magnetic field on a field emission current in logarithmic coordinates according to the generalized Fowler-Nordheim equation. Solid line shows the field emission current in the absence of a magnetic field. The current in the presence of an external magnetic field of 0.5 T parallel to the surface is shown with the dotted line.}
\label{ris:image}
\end{figure}

\section{Estimates of the magnetic field influence}
Experiments at CERN are planned on the setup with small interelectrode gap. In this case field emission current plays a major role in breakdown occurrence. Therefore, it is necessary to calculate the influence of a magnetic field on the field emission current. Taking into account that the electric field can increase on the asperities by 30-140 times \citep{Descoeudres} the real value of the electric field is: $E'=\beta E$, where $\beta$ is an enhancement factor. Let's choose the values of electric and magnetic fields from the experiments at CERN $E=144 MV/m$, $B=0.5 T$ and an enhancement factor $\beta=30$. The influence of a magnetic field on the field emission current will be:
\begin{equation}
\left|\vec{j_B}\right|=j_{FN}\cdot\left(e^{-\frac{b}{\beta E}}\right)^{\frac{1}{2\beta^2}}\left(1-\frac{1}{2\beta^2}\right)\approx j_{FN}\cdot\left(1-5\cdot10^{-4}\right)
\end{equation}
It can be seen that in practice the relatively small values of the magnetic field induction do not affect the field emission current. And we can predict that effect of presence of a magnetic field will be hardly observable. Obviously, the effect of such a magnetic field on the current of emitted ions will not be observable at all.

Experiments which are planned at IAP NASU on the setup with large interelectrode gap should depend of processes in the interelectrode gap \citep{Tarasova, Seznec}. In \citep{Mazanko} authors shows that a varying magnetic field (even a small value $0.03~T$) enhance diffusion and phase formation and increase $j_{diff}^{(0)}$. Other neutral flows $\left(j_{rest}^{(0)}, j_{desor}^{(0)}, j_{T,evap}^{(0)}, j_{Ion,atom}^{(0)}\right)$ not dependent on the magnetic field. However, their ionization process leads to the appearance fluxes of charged particles which will be affected by the magnetic field. Therefore, it is necessary to consider the influence of the magnetic field on ionization currents in the interelectrode gap. As follows from \citep{Engel}, the electron energy range of $10-100 eV$ is the most effective for ionization. For the electric field strength $E =100 MV/m$ it corresponds to a layer of $0.1-1\mu m$. In this region cathode plasma is formed. Electrons emitted from a metal or formed in the inter-electrode gap quickly gather the energy more than the effective. Therefore, they will not ionize the neutral atoms. But an external magnetic field can cause lengthening of the electron trajectory (especially for electrons with energies not greater than an effective energy). Therefore it can influence on the process of ionization.

Let's evaluate the relative elongation of the electron trajectory in presence of magnetic field. It is well known that an electron in mutually perpendicular electric and magnetic fields moves along trochoidal trajectory \citep{Landau}. Let's write the parametric equations of trajectory of an electron in this configuration of fields in dimensionless coordinates:
\begin{equation}
\begin{array}{l}
\xi=\left(1+\tilde{v}_{0x}\right)\sin{\tau}+\tilde{v}_{0y}(\cos{\tau}-1)-\tau \\
\zeta=\tilde{v}_{0y}sin{\tau}+(1+\tilde{v}_{0x})(1-\cos{\tau})
\end{array}
\end{equation}
where $t_0={2\pi}/{\omega_B}$, $L_{dr}=ct_0$, $\xi=x/L_{dr}$, $\zeta=y/L_{dr}$, $\tau=t/t_0$, $\tilde{v}_{0x}=v_{0x}/c$, $\tilde{v}_{0y}=v_{0y}/c$, $\omega_B$ is a gyrofrequency, $v_{0x}$ and $v_{0y}$ are the initial velocities along corresponding axes.

The length of the trajectory can be found by the formula:
\begin{equation}
l=\int_0^{\tau_{h_d}}\sqrt{v^2_x+v^2_y}d\tau\approx h_d+\tilde{v_0}\sqrt{2h_d},
\end{equation}
where $h_d=h|_{y=d}=d/L_{dr}$.
Using values from IAP NASU experiment ($B=0.33~T,~E=100~MV/m$ and $d=100~ \mu m$) we can evaluate parameters of (20):
$t_0=10^{-10}~s$, $L_{dr}=3.25~cm$ and $h_d=0.003$

The magnetic field substantially changes the trajectory of ionization electrons with the energy of several eV directed to the cathode. The maximum of ionization occurs for electrons with energy of tens of eV \citep{Engel}. From this, it follows (taking into account ionization energy) that electrons which form in the gap have energies of several eV. As an example electron with energy $\varepsilon=2 eV$ have been chosen.
Considering that velocity is $v_0=\sqrt{\frac{2\varepsilon}{m}}$ the relative elongation of the electron trajectory will be:

\begin{equation}
\frac{\Delta l}{l}=\tilde{v_0}\sqrt{\frac{2}{h_d}}\approx 0.1
\end{equation}

It should be emphasized that the elongation of the electron trajectory leads to increasing of the electron current due to increasing of the ionization. In the case of the (cut-off magnetic field $B_H$) all of the formed in the interelectrode gap electrons (both emitted and ionized) do not reach the anode. And this case the elongation of the electron trajectory has no influence on the breakdown probability.

In addition, the magnetic field changes the angle of entry of the electron to the metal and this may affect on the desorption process. Estimates of the electron entrance angle to the surface of the anode gives: $\theta\approx\frac{v_x}{v_y}=\sqrt{\frac{h_d}{2}}=2^o$. The inclination is small and the effect of increasing of the dark current will be weak.

Summarizing the foregoing the magnetic field can affect on the: field emission current density $j_B$, angle of entry of electrons to the surface of the anode, thereby affecting on desorption, diffusion of gases from the volume of electrodes. We predict that the most visible effect causes by increasing of the ionization as a result of elongation of the electrons trajectories under static magnetic field influence.

\section{Experiments at CERN}
\begin{figure}[h]
\center{\includegraphics[width=1\textwidth]{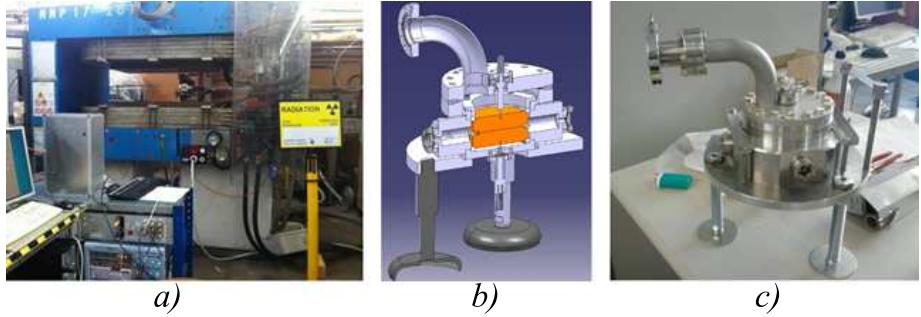}}
\caption{The facility for studying high-gradient DC breakdowns in the magnetic field at CERN: a) LES inside of electromagnet, b) 3D model of LES, c) the photo of LES.}
\label{ris:image}
\end{figure}

A Large Electrodes System (LES) is a compact vacuum system containing two electrodes with a relatively large surface area for pulsed DC system. The LES was especially designed to be small enough to fit inside a large aperture $0.5~T$ dipole electromagnet (Fig. 2). The vacuum chamber is made of stainless steel and has four quartz windows located symmetrically opposite one another. The chamber is connected to a vacuum system that provides vacuum of order $10^{-7} Pa$.  Diameter of electrodes is $62 mm$ and roughness is less than $1 \mu m$ (Fig.~3). The fabrication steps for CLIC accelerating structures, so what we copy for the electrodes, is given in \citep{Wang}. Was used the standard cleaning procedure for Cu at CERN \citep{Malabaila}. The distance between the electrodes is regulated by the choice of ceramic inserts (rings) between the electrodes.

\begin{figure}[h]
\center{\includegraphics[width=0.7\textwidth]{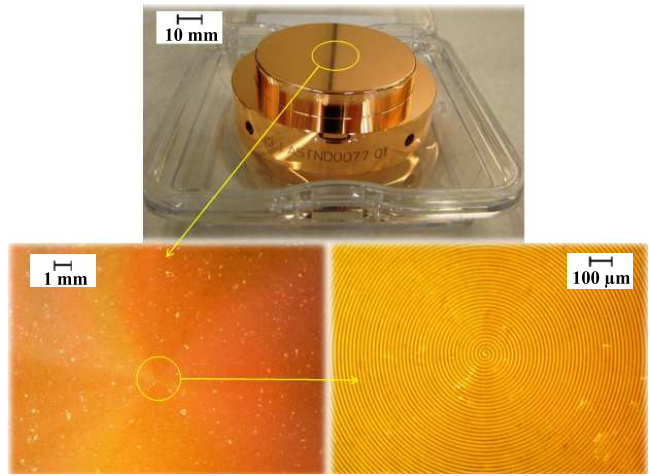}}
\caption{The images of the copper electrode.}
\label{ris:image}
\end{figure}

The available electrical circuit allows application of $\vec{E}$ with frequency $1$ $kHz$. A detailed description of the electric circuit and the principle of operation can be found in \citep{Shipman,Shipman2}. The value of applied voltage was selected manually to ensure the breakdown rate $\left(BDR=\frac{number \ of \ breakdowns}{number \ of \ pulses}\right)$ $10^{-4} \div 10^{-5}$.

The influence of a magnetic field on the BDR was tested in the Large Electrodes System \citep{Shipman,Shipman3} with an electrode gap of $15 \mu m$ for a magnetic field parallel to the surface of the electrodes. A magnetic field strength was set to $0.5~T$ and the voltage was set to $2.16$ $kV$ that corresponds to electric field strength  $144 MV/m$ (taking no account of an enhancement factor).BDR should be measured over long period of time to detect a potentially small change due to magnetic field. The electrodes of the LES were known to condition with time. Therefore, to prevent  masking the effect of the magnetic field on the BDR  by the conditioning of the electrodes, the BDR was measured over relatively brief periods of time before the field of the magnet was changed and a new BDR measurement was started. The experimental procedure was as follows: the LES was first installed in the center of the magnet; with the switch initially off the electric field was varied until a BDR of approximately $10^{-4}$ was obtained; a set number of BDs (usually $20$ or $30$) were obtained and the BDR recorded; after the set number of BDs had been reached the switch was stopped from pulsing; the magnet was then switched on and ramped up to full power over the cause of a couple of minutes; once the set number of BDs had been recorded with the magnet switched on and the BDR over this period recorded, the switch was again stopped. This process was then repeated as often as possible.
\begin{figure}[h]
\center{\includegraphics[width=0.9\textwidth]{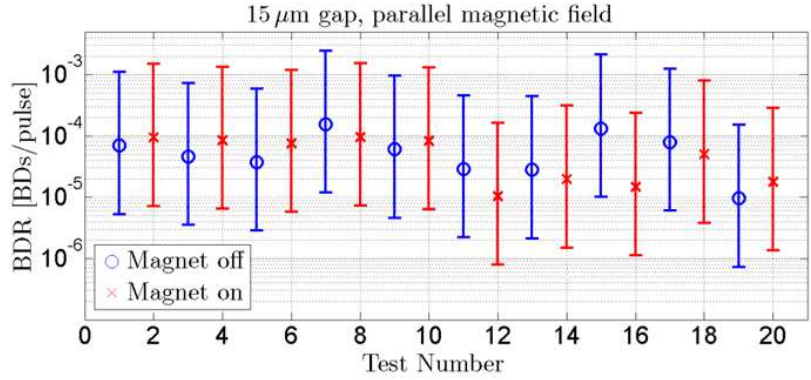}}
\caption{The measured BDR in the Large Electrodes System with a $15 \ \mu m$ gap size with and without an external about $0.5 \ T$ magnetic field applied parallel to the electrodes.}
\label{ris:image}
\end{figure}

Fig. 4 shows BDR in a set of measurements with a magnetic field parallel to the surface of the electrodes. As can be seen there is no average discernible difference between the BDR with or without a magnetic field regardless of the field orientation. Though some of the data recorded seems to suggest a small increase in the BDR due to the presence of a magnetic field.

\section{EXPERIMENTS AT IAP NASU}
At the IAP NASU the study of the influence of an external magnetic field on the field emission current and the breakdown voltage was carried out on the experimental facility described in \citep{Baturin}. The block diagram of the installation is shown in Fig. 5.

The experimental facility includes the following elements: a vacuum chamber with samples and a monolithic RGA mass spectrometer (Residual Gas Analyzer) for analyzing the residual gas in a vacuum chamber; precision anode-cathode positioning device (manipulator); high-voltage power supply; dark current and breakdown voltage measurement system.
\begin{figure}[h]
\center{\includegraphics[width=0.5\textwidth]{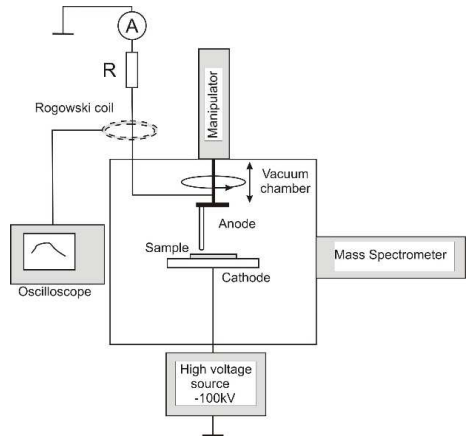}}
\caption{ Block diagram of the installation for the study of high-voltage breakdowns}
\label{ris:image}
\end{figure}

The test samples are placed in a working chamber, which is pumped down to pressure $\sim 10^{-7} \ Pa$. Electrodes of the discharge gap have a so-called "tip-plane" configuration. The anode is a rod of a diameter of $2.5 \ mm$ which ends with a rounded part in the form of a hemisphere. The cathode is a disk of diameter $12 \ mm$ and thickness $2 \ mm$. The test sample is located on the holder, which is the cathode. Specially developed vacuum rotary-translatory input $(manipulator)$ allows precise adjustment of the distance between electrodes in a wide range with an accuracy of $5~\mu m$.  It's possible to move the anode over the cathode along a circle with a radius of $4 \ mm$ with an accuracy of $7$ degrees. The dark current and the breakdown voltage are registered in the anode circuit. A negative DC voltage in the range $1 \div 100 \ kV$ from a high-voltage power supply is applied to the cathode.

A magnetic system based on Sm-Co magnets was used in the experimental facility. Cone-shaped concentric tips were used for the magnetic field amplification on the axis of the magnetic system. The view of magnetic system is shown in Fig.6. Magnetic field in the region of the discharge gap is $0.33 \ T$. The test sample was located between the pole pieces and the magnetic field was oriented parallel to the electrode surface.

\begin{figure}[h]
\center{\includegraphics[width=0.3\textwidth]{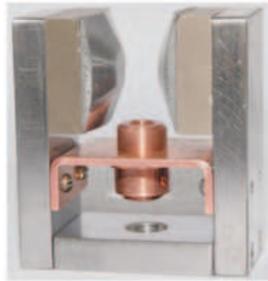}}
\caption{The construction of the magnetic system}
\label{ris:image}
\end{figure}

The test samples (cathodes) were made of copper with a low content of impurities. Surface of samples was mechanical ground and polished. A part of the samples was ion-purified in the glow discharge plasma in the vacuum chamber of the setup. Some samples were exposed to chemical etching with different etching times: 1, 3 and 5 minutes. Copper anode was used in the experiments.

The effect of the magnetic field on the BDR is expected to be very small and at the limit of resolution in the  CERN set up. The distance between the electrodes would have to be increased to make effect stronger. As a consequence, the maximum possible inter-electrode gap of $100~ \mu m$ was used to measure the effect in the  experiments. The values of the high voltage were gradually increased creating a constant field in the discharge gap with a voltage of $1  \ MV/m$ and higher up to the breakdown. The values of the dark current between the cathode and the anode were measured in the process of increasing the voltage at the cathode. Typically, short-term, low-power, self-extinguishing breakdowns arising in the process of increasing the field strength between the electrodes don't lead to change in the insulation strength of the discharge gap and contribute to the cleaning of the surface of the electrodes The high-voltage breakdown leads to the formation of a high-current spark discharge evolving into an arc discharge in the metal vapor of the electrodes.
\begin{figure}[h]
\center{\includegraphics[width=0.8\textwidth]{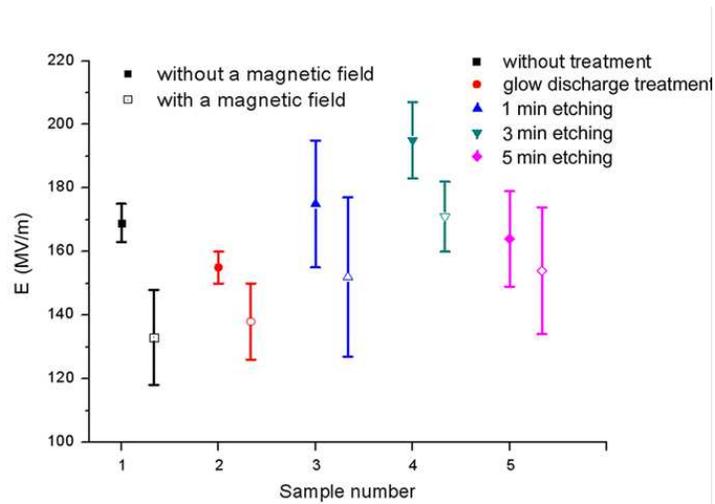}}
\caption{The dependence of breakdown voltages of different surface treatments for copper samples in the presence and absence of a magnetic field.}
\label{ris:image}
\end{figure}
The values of the dark current and the breakdown voltages were measured at several anode-cathode positions and then averaged for each sample. Measurements of the breakdown voltage and the dependence of the dark current on the electric field strength were carried out in the presence of a magnetic field and without it for each of the samples under study.  Figure 7 shows a histogram of the breakdown voltage $E$ for a series of samples with different surface treatments.
\begin{figure}[h]
\center{\includegraphics[width=1\textwidth]{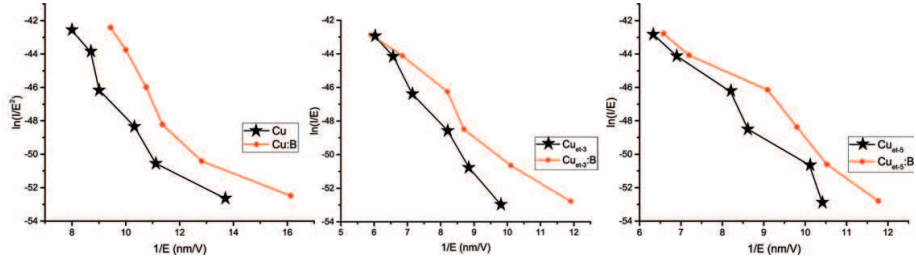}}
\caption{Dependence of the dark current on the electric field strength}
\label{ris:image}
\end{figure}
As can be seen from the figure, the application of a magnetic field to the discharge gap reduces the average breakdown voltage by $(10 \div 20) \% $ for all samples regardless of the way the surface of the cathode is treated.

The dependences of the values of dark currents on the electric field strength for a series of cathode samples with different surface treatments in the presence of a magnetic field parallel to surface and without a magnetic field are shown in Fig. 8. The following notations are used in the figure: $Cu$ is the cathode outside magnetic field; $Cu:B$ is the cathode in a magnetic field; $Cu_{d}$ is the cathode that was treated in a glow discharge; Cu$_{et-1}$, Cu$_{et-3}$,  Cu$_{et-5}$ are cathodes that were etched 1, 3 and 5 minutes.

It can be seen from the figure that the value of the dark current increases monotonically with increasing electric field strength for all cathode samples and the application of a magnetic field leads to its insignificant growth.

\section{Conclusions}
Theoretical treatment is given to diverse particle flows found in the interelectrode gap and the possibility of controlling them by means of an external magnetic field parallel to the surface. Analysis was performed of all the charged- and neutral flows contributing to the breakdown occurrence. In the case of small interelectrode gaps usually the major process is field emission of electrons. By "switching-off" the field emission current by means of the magnetic field, $B_H$, parallel to the metal surface, it is possible to inhibit breakdowns. With the accelerating gradient target of $100~ MV/m$ for the interelectrode gap of $100~ \mu m$ the "switching-off" field $B_H=3.4~ T$ and for the $10~ \mu m$ gap $B_H=10.7~ T$. This way has the technical complications and can lead to degradation of the acceleration properties. For large gaps the general theory is not present now. But it is known that the electron and ion bombardment of electrodes usually plays major role in the breakdowns formation. The ion current was estimated to be 3 orders of magnitude smaller than the electron current. Yet, bombarding the electrode by more energetic ions leads to significant damage provoking a subsequent avalanche-type current increase.

This paper presents a generalization of the Fowler-Nordheim equation, enabling us to take into account the effect of a magnetic field parallel to the surface on the density of the field emission current.
For typical experimental values of $E$ and $B$, the effect of the magnetic field on the emission current is about $0.05\%$. Needless to say this magnetic field effect on heavy ions may well go unnoticed.
The effect of the magnetic field on the metal vapour- and gas ionizations results in the production of electron flows ranging in energy from $1$ to $10~eV$. For the electric field intensity $E=100~MV/m$ the ionized layer is $0.1$ to $1~\mu m$. In \citep{Shipman3} it is shown that the plasma layer thickness is determined by a number of factors including a voltage drop, decreased electron density and temperature, etc. Electrons of these energies heading for the cathode change their trajectories under the action of the magnetic field parallel to the metal surface. In the presence of the external magnetic field $B=0.3 \div 0.5 ~T$ the length of the low-energy electron path is increased by tens percent and due to the increment in electron-impact ionization. A magnetic field modifies the angle of electron entrance into the metal, which might influence the desorption. Estimates made for similar fields reveal approximately a $2^o$ change in the incidence angle of electrons reaching the anode surface. The modified angle of incidence leads to the increased desorption, but for these values of the field the effect would not be significant. Thus, the most pronounced effect exerted by the magnetic field is the longer electron paths, and thus, higher ionization probability.

Experiments to examine the magnetic field effect on BDR were done at CERN $(E=144~ MV/m,~ B \sim 0.5 ~T,~d=15~\mu m)$ and  on the field emission current $(E=60 \div 160~ MV/m, B \sim 0.3~ T,~d=100~\mu m )$ at the Institute of Applied Physics, National Academy of Sciences of Ukraine (IAP NASU).
Experiments at CERN was made on the setup with small interelectrode gap. In this case field emission current plays a major role in breakdown occurrence. No appreciable difference was found between BDR with a magnetic field and BDR without it.  But some recorded data seem to indicate a slight increase  in BDR in the presence of a magnetic field. This results are consistent with theoretical prediction of the magnetic field influence.
Moreover, experiments made at IAP NASU on the setup with large interelectrode gap shows (Fig.7-8)  that with a magnetic field present the dark current between electrodes is increased and the breakdown voltage is decreased by $(10-20)\%$ for practically all the cathodes no matter what their surface treatment. As it was theoretically predicted due to increasment in electron-impact ionization.
Although these results are in satisfactory agreement with theoretical predictions, yet further thorough work, both theoretical and experimental, is needed.

\section{Acknowledgments}
This work was supported by  the target research program of the NAS of Ukraine cooperation with CERN and JINR "Nuclear matter in extreme conditions" under  Contract CO-1-16/2017

\end{document}